\documentclass[11pt,twoside]{article}
\usepackage{asp2014}

\aspSuppressVolSlug
\resetcounters

\begin{document}

\title{Automated SpectroPhotometric Image REDuction (ASPIRED)}

\author{Marco~C~Lam, Robert~J~Smith, and Iain~A~Steele}
\affil{Astrophysics Research Institute, Liverpool John Moores University, IC2, LSP, 146 Brownlow Hill, Liverpool, Merseyside L3 5RF, UK\\
\email{c.y.lam@ljmu.ac.uk}}

\paperauthor{Marco~C~Lam}{C.Y.LAM@ljmu.ac.uk}{0000-0002-9347-2298}{Astrophysics Research Institute}{Liverpool John Moores University}{Liverpool}{Merseyside}{L3 5RF}{UK}
\paperauthor{Robert~J~Smith}{R.J.SMITH@ljmu.ac.uk}{}{Astrophysics Research Institute}{Liverpool John Moores University}{Liverpool}{Merseyside}{L3 5RF}{UK}
\paperauthor{Iain~A~Steele}{I.A.STEELE@ljmu.ac.uk}{}{Astrophysics Research Institute}{Liverpool John Moores University}{Liverpool}{Merseyside}{L3 5RF}{UK}



  
\begin{abstract}
We aim to provide a suite of publicly
available spectral data reduction software to facilitate
rapid scientific outcomes from time-domain observations. For time 
resolved observations, an automated pipeline frees astronomers from
performance of the routine data analysis tasks to concentrate on
interpretation, planning future observations and communication with
international collaborators. The
project~consists of two parts: data processing~(\texttt{ASPIRED})
and a graphical user interface~(\texttt{gASPIRED}).
\texttt{ASPIRED} is written in \texttt{Python~3}, and was intentionally developed
as a self-consistent reduction pipeline with its own diagnostics and
error handling. The pipeline can reduce 2D spectral data
from raw image to wavelength and flux calibrated 1D spectrum
automatically without any user input. \texttt{gASPIRED} is a
cross-platform software developed with \texttt{Electron} on a single
code base. It brings interactivity to the software with a well-maintained
and user-friendly environment.
\end{abstract}

\section{Introduction}
With major global investments in multi-wavelength and multi-messenger surveys, time domain astronomy is entering a golden age. To maximally scientific exploit
discoveries from these facilities rapid spectroscopic follow-up observations of transient
objects~(e.g.,\ supernovae, gravitational wave optical counterparts etc.)
will provide crucial {\em astrophysical}
interpretations. Part of the OPTICON\footnote{\url{https://www.astro-opticon.org/}} project
coordinates the operation of a network of
largely self-funded European robotic and conventional telescopes, coordinating
common science goals and providing the tools to deliver science-ready photometric and spectroscopic data.  As part of this activity SPRAT ~\citep{2014SPIE.9147E..8HP} was developed as a compact, reliable, low-cost and high-throughput spectrograph and appropriate for deployment on a wide range of 1-4m class telescopes.
Installed on the Liverpool Telescope since 2014, the deployable slit
and grating mechanism and optical fibre based calibration
system make the instrument self-contained.  Copies of SPRAT are being built for other telescopes.  Our final task
is to deliver software that can be easily configured to build
pipelines for long slit spectrographs on different telescopes.   We use SPRAT as a test case for this development, but our aim is to support a much wider range of instruments.  By delivering near real-time data reduction we will facilitate automated or interactive decision making,
allowing "on-the-fly" modification of observing strategies and rapid triggering of other facilities.

\section{Desktop or Web Application?}
Desktop and web-based applications have their strengths and
weaknesses, examples specific to the development of the ASPIRED
pipeline are summarised in Table~\ref{tab:comparison}. Some other
considerations include:\\
\textbf{Geographical Location} -- If users are working in
extreme or remote environments, there is less likely to be a high speed and
reliable internet connection. An application will become completely
unusable if it does not have an offline mode.\\
\textbf{Platform} -- The three platforms that dominate the
market are the various distributions of Windows, Mac and Linux.
(In the context of research data reduction, we neglect hand-held
and mobile computing. The solutions selected here ought to be well
suited to adaptation to the current generation of mobile devices.)
Astrophysics research heavily employs Unix-like systems,
but Windows is more commonly used in teaching. \\
\textbf{Cost} -- Development costs for a web-based and a desktop
application are similar. When the technology shifts to a completely
new paradigm, updating a piece of software can be more expensive than
developing a new one.

\renewcommand{\arraystretch}{1.2}
\begin{table}
    \centering
    \begin{tabular}{c|c}
        \textbf{Desktop} & \textbf{Web-based} \\
        \hline
        Users only need to download once & Has to be downloaded every time\\
        Users have to update the service & A single copy is maintained on the server\\
        Work both on- and/or off-line & Only work online\\
        No overhead cost for hosting & Cost for hosting service\\
        Difficult cross-platform support & Difficult cross-browser support\\
        High performance for local storage & Usually up/download speed limited\\
        Less concern with security problem & Every connection is a potential security issue\\
        
    \end{tabular}
    \caption{Comparison of the pros and cons of desktop and web-based application.}
    \label{tab:comparison}
\end{table}

\section{Data Processing Software}
In order to maximise the potential user base and allow easier
maintenance and future extension, the development is divided
into three self-consistent components -- (a)~2D and 1D spectral
reduction~(except wavelength calibration); (b)~wavelength
calibration~(see \texttt{RASCAL} by \citet{P10-37_adassxxix} in
this proceeding); and (c)~graphical user interface. The 
\texttt{ASPIRED} is written in object-oriented \texttt{Python~3},
while the \texttt{gASPIRED} is written in \texttt{JavaScript},
\texttt{html} and \texttt{CSS} with the
\texttt{Electron}\footnote{\url{https://electronjs.org/}} framework.

\subsection{Data Reduction -- \texttt{ASPIRED}}
\begin{figure}
    \centering
    \begin{tabular}{cc}
        \begin{minipage}{0.46\textwidth}
            \includegraphics[width=\textwidth]{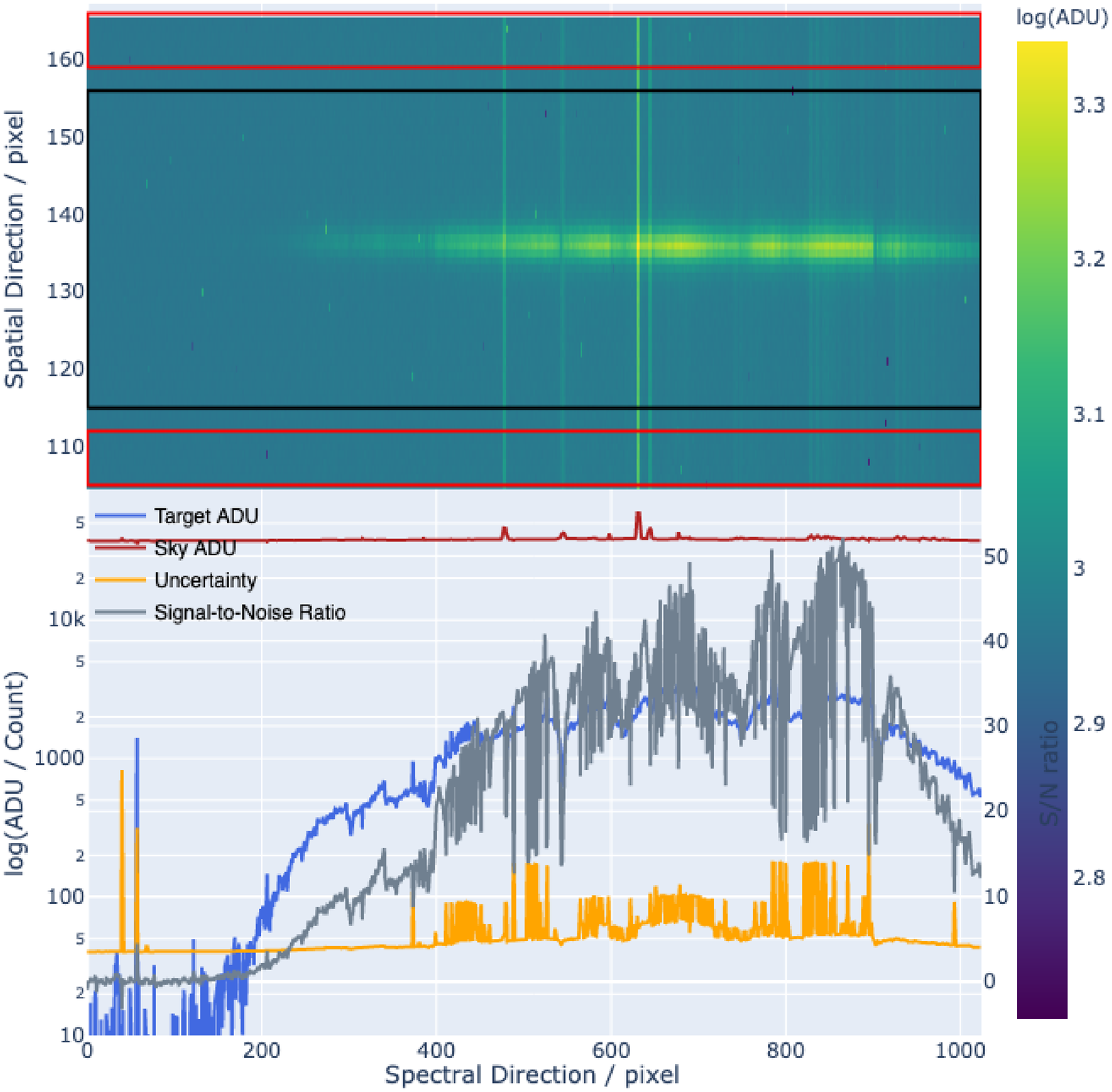} 
        \end{minipage}&
        
        \begin{minipage}{0.46\textwidth} 
            \includegraphics[width=0.98\textwidth]{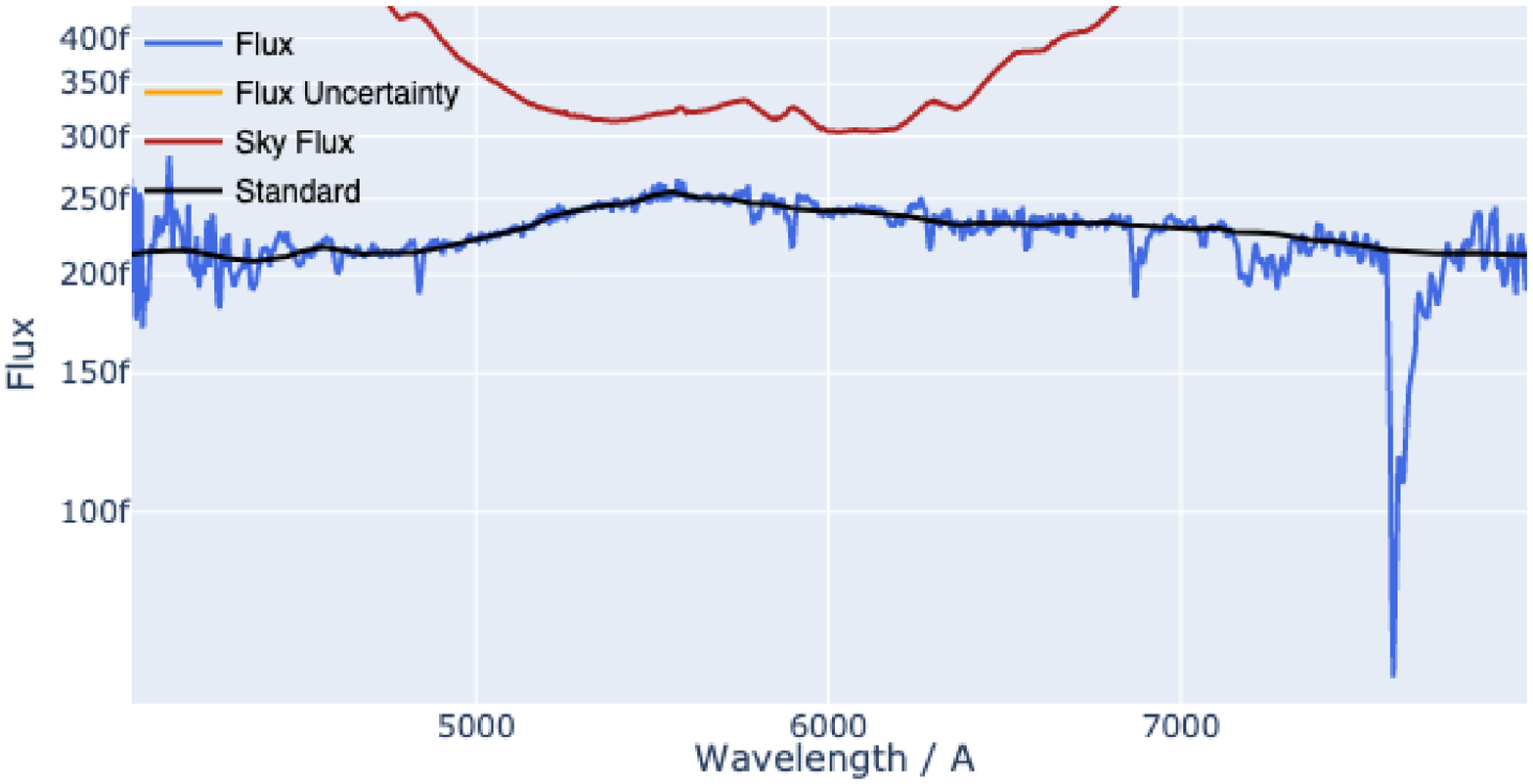} \\ 
            \includegraphics[width=\textwidth]{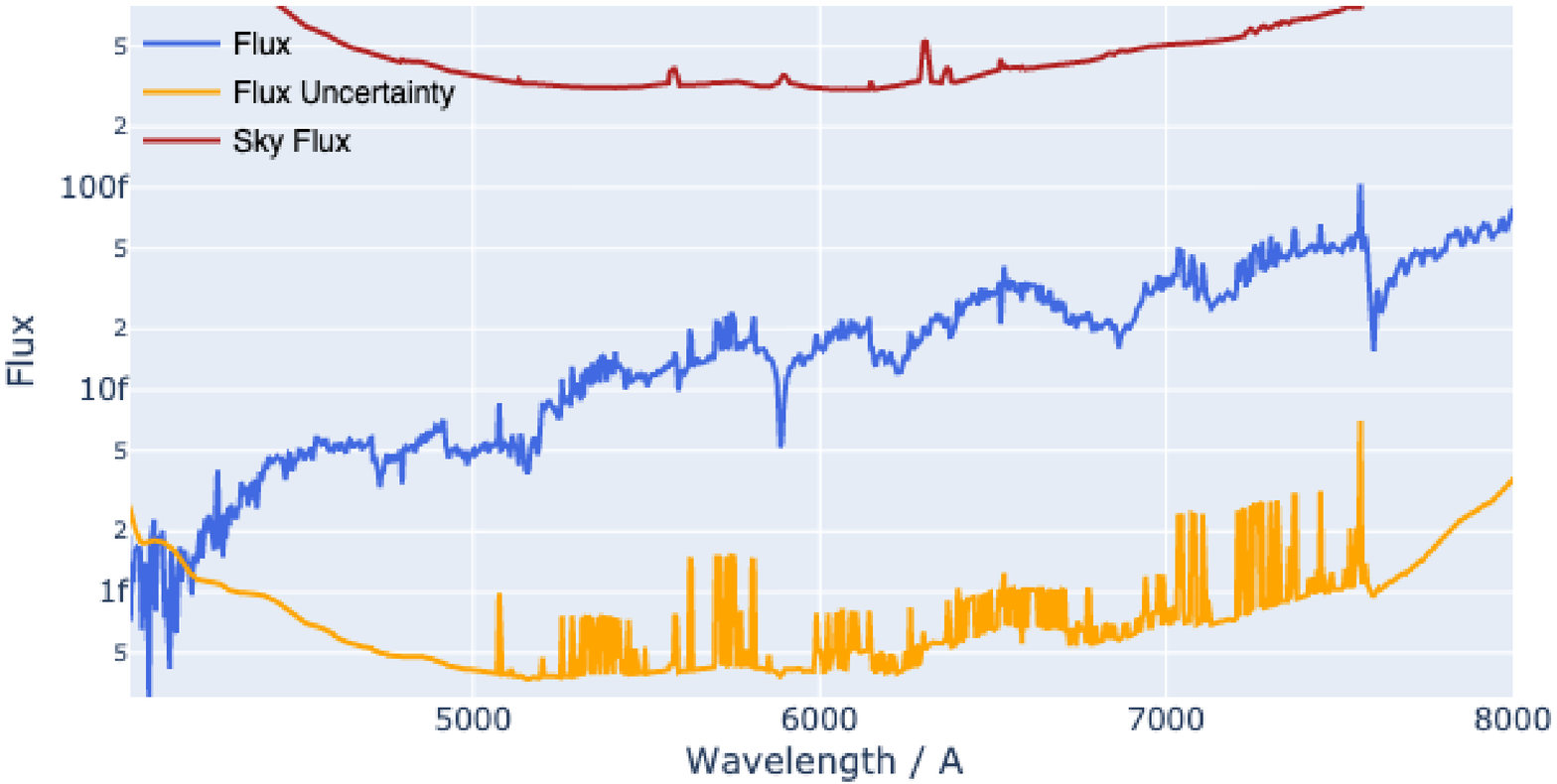} 
        \end{minipage}
    \end{tabular}
    \caption{An M-dwarf spectrum taken with SPRAT in red optimised mode
    used as a test case for the development. Top Left: visualisation of
    the regions for spectral extraction region~(black) and background
    sky fitting~(red); bottom left: extracted spectrum~(blue),
    uncertainty~(orange), sky signal~(red) and signal-to-noise
    ratio~(grey); top right: the extracted spectrum of the standard
    star after wavelength and flux calibration~(blue) and the standard
    spectrum used for flux calibration~(black); bottom right: the
    extracted spectrum~(blue), the uncertainty~(orange) and the sky
    signal~(red).}
    \label{fig:visulation}   
\end{figure}

\texttt{ASPIRED}\footnote{\url{https://github.com/cylammarco/ASPIRED}} focuses on spectral data extraction, but it also 
contains a minimal image reduction facility, which serves basic
arithmetic operations. The data extraction itself contains:
(a) aperture tracing; (b) top hat/optimal aperture
extraction~\citep{1986PASP...98..609H};
(c) wavelength calibration with \texttt{RASCAL}, or user defined
polynomial/spline solutions; and (d) flux calibration if standard
star is also provided. It is currently suitable for low resolution
long-slit spectral extraction, including flux calibration.
It makes use of a number of well-developed
packages: \texttt{NumPy}\citep{Walt:2011:NAS:1957373.1957466}, \texttt{ccdproc}\citep{matt_craig_2017_1069648} \& \texttt{Astropy} for
data handling~\citep{2013A&A...558A..33A, 2018AJ....156..123A};
\texttt{SciPy}\citep{2019arXiv190710121V} for peak finding and curve fitting, as well as for
a 1D geometrical distortion correction along the spatial direction,
which is applicable to cases of single or multiple sources appearing in
the slit, it should
be easily extendable to spectral images taken with most integrated field
units~(although it currently does not support flux calibration of
this type of spectrograph); \texttt{Astro-SCRAPPY}\citep{curtis_mccully_2018_1482019, 2001PASP..113.1420V} for cosmic ray
removal to improve the reliability in spectral tracing~(original
image is used in spectral extraction); \texttt{RASCAL} for wavelength
calibration; a slightly modified forked version of
\texttt{SpectRes}\footnote{\url{https://github.com/cylammarco/spectres}}
for 1D spectral resampling~\citep{2017arXiv170505165C}; and
\texttt{plotly}\citep{plotly} for visualisations~( Fig.~\ref{fig:visulation}) in
both scripting mode and passing JSON to \texttt{gASPIRED} for the
interactive GUI mode.

\subsection{Graphical User Interface -- \texttt{gASPIRED}}
As a layer independent of the ``logical'' part, \texttt{gASPIRED}
itself cannot function without an appropriate data reduction script
running at the back of the GUI. The spectral identification from a
2D image is manipulated with JS9\footnote{\url{https://js9.si.edu}}
~\citep{eric_mandel_2019_3382228}, JSON strings are sent to the back
calling a python script to perform the spectral identification and
extraction interactively. The extracted spectra are displayed using
\texttt{plotly}. Because the \texttt{Python-plotly} is built
on top of the \texttt{JavaScript} version, it allows us to maintain
only a single code base for visualisation in both scripting and GUI
modes. If developers prefer a different plotting library, or a static
image file is preferred over dynamic plots, only minor modification
would be required. The \texttt{Electron} framework has cross-platform
support. \texttt{Travis CI}
is used to test the cross-platform builds from Linux and Mac environments 
for deployment to Windows~(32-bit), Linux and Mac. The final software can be
installed or used as a ``double-click'' portable application. This is
also hosted on GitHub\footnote{\url{https://github.com/cylammarco/gASPIRED}}
in a public repository.

\section{Future Development}
The current development is expected to be released as an alpha version
at the end of 2019. 
We anticipate wider beta-release in 2020.  The object-oriented
nature of the development should make it easy for subseqeuent conversion into an
\texttt{Astropy} affiliated package should it become the future goal.

\section*{Acknowledgement}

This project has received funding from the European Union's Horizon 2020 research and innovation programme under grant agreement No 730890. This material reflects only the authors views and the Commission is not liable for any use that may be made of the information contained therein.







\bibliography{P10-46.bib}


\end{document}